\documentclass[journal]{vgtc}                
\ifpdf
  \pdfoutput=1\relax                   
  \usepackage{graphicx}                
  \DeclareGraphicsExtensions{.pdf,.png,.jpg,.jpeg} 
\else
  \usepackage{graphicx}                
  \DeclareGraphicsExtensions{.eps}     
\fi%

\graphicspath{{figures/}{pictures/}{images/}{./}} 

\usepackage{xcolor}
\usepackage{microtype}                 
\PassOptionsToPackage{warn}{textcomp}  
\usepackage{textcomp}                  
\usepackage{mathptmx}                  
\usepackage{times}                     
\usepackage{cite}                      
\usepackage{enumitem}
\usepackage{tabu}                      
\usepackage{booktabs}                  
\usepackage{setspace}
\usepackage{doi}
\RequirePackage{doi}
\usepackage{hyperref}
\usepackage[linesnumbered,ruled,vlined]{algorithm2e}
\usepackage{amsmath}

\hypersetup{
    colorlinks,
    linkcolor={red!50!black},
    citecolor={blue!50!black},
    urlcolor={blue!80!black}
}

\newcommand{\cP}[1]{\textcolor[rgb]{.957,.478,.271}{#1}}
\newcommand{\cO}[1]{\textcolor[rgb]{.306,.4,.502}{#1}}
\newcommand{\cW}[1]{\textcolor[rgb]{.396,.824,.447}{#1}}
\newcommand{\cL}[1]{\textcolor[rgb]{.867,.376,.376}{#1}}
\newcommand{\ec}[1]{``\textit{#1}''}
\newcommand{\revise}[3]{#3}

\newcommand{\purplearrow}{\textcolor[rgb]{.4,.176,.569}{purple arrows}}
\newcommand{\greenarrow}{\textcolor[rgb]{0,.408,.216}{green arrows}}
\newcommand{\redarrow}{\textcolor[rgb]{1,0,0}{red arrows}}
\newcommand{\purpleArrow}{\textcolor[rgb]{.4,.176,.569}{Purple arrows}}
\newcommand{\greenArrow}{\textcolor[rgb]{0,.408,.216}{Green arrows}}
\newcommand{\redArrow}{\textcolor[rgb]{1,0,0}{Red arrows}}

\SetKwInput{KwInput}{Input}                
\SetKwInput{KwOutput}{Output}              

\onlineid{1041}

\vgtccategory{Research}
\vgtcpapertype{system}

\title{RASIPAM: Interactive Pattern Mining of\\Multivariate Event Sequences in Racket Sports}

\author{Jiang Wu, Dongyu Liu, Ziyang Guo, and Yingcai Wu}
\authorfooter{
\item
 J. Wu, Z. Guo, and Y. Wu are with the State Key Lab of CAD\&CG, Zhejiang University. Y. Wu is also with the Alibaba-Zhejiang University Joint Research Institute of Frontier Technologies. E-mail: \{wujiang5521, ziyangguo27, ycwu\}@zju.edu.cn. Y. Wu is the corresponding author.
\item
 D. Liu is with MIT. E-mail: dongyu@mit.edu.
}

\shortauthortitle{Wu \MakeLowercase{\textit{et al.}}: RASIPAM: Interactive Pattern Mining of Multivariate Event sequences in Racket Sports}

\teaser{
  \centering
  \includegraphics[width=\linewidth]{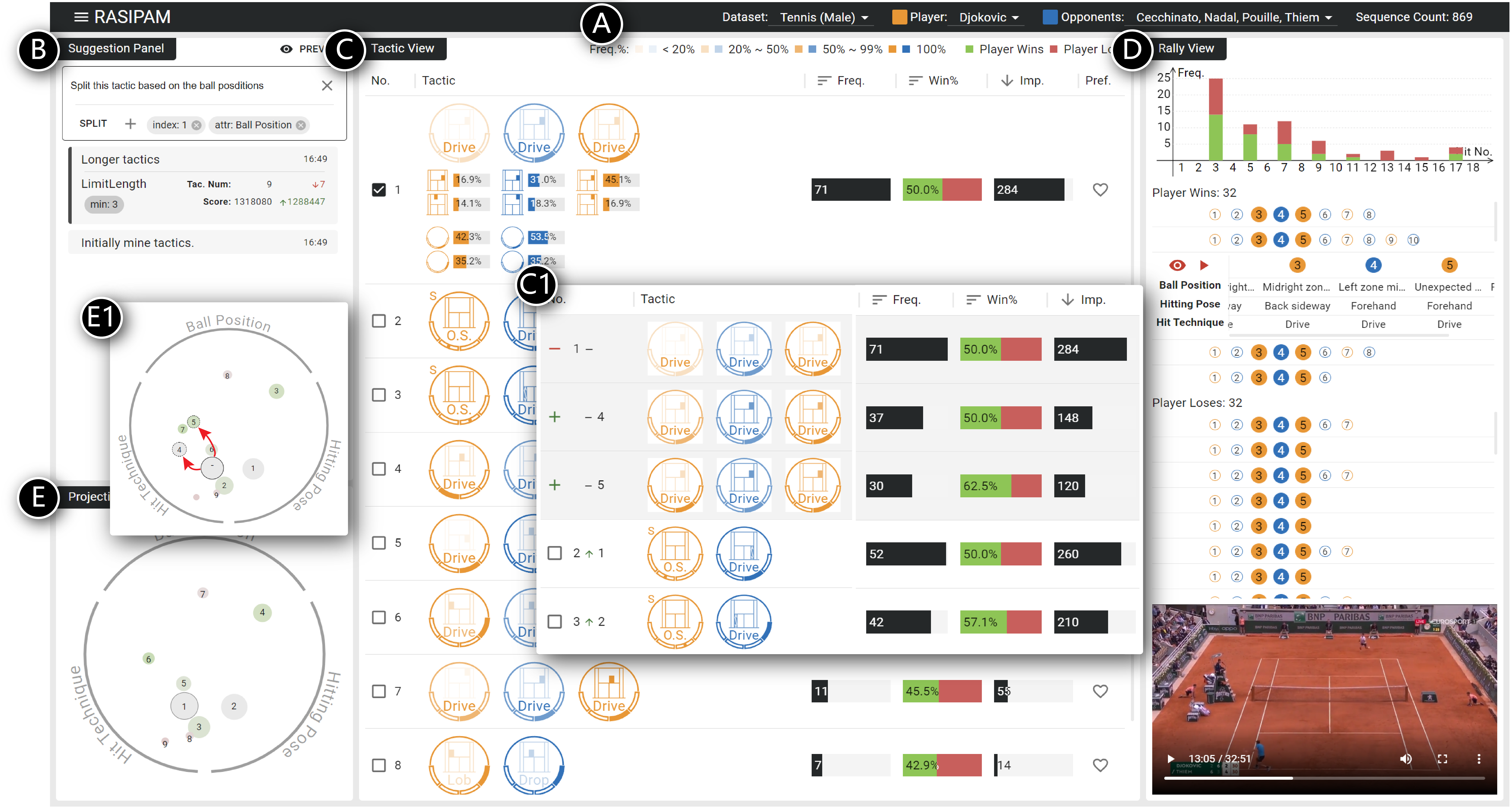}
  \caption{
    A screenshot of our system for interactive tactic mining.
    Domain experts first need to select a player of interest on the control bar (A) to analyze his/her tactics.
    Then the system mines a initial set of tactics and visualizes them through intuitive glyphs in the Tactic View (C).
    Meanwhile, the Project View (E) provides an overview of the tactics through a semantic projection method, allowing domain experts to quickly understand the tactics.
    Experts are then allowed to give suggestions to improve the tactics based on their domain knowledge in the Suggestion Panel (B).
    The system will refine the tactic set and compare the new tactics with the original ones (C1 and E1).
    For detailed exploration, the Rally View (D) shows more details about a chosen tactic, including a bar chart displaying where the tactic is used, a list of all rallies involving the tactic, and a video player for the video segment of each rally.
  }
  \label{fig:teaser}
}

\abstract{
    Experts in racket sports like tennis and badminton use tactical analysis to gain insight into competitors' playing styles. Many data-driven methods apply pattern mining to racket sports data — which is often recorded as multivariate event sequences — to uncover sports tactics. However, tactics obtained in this way are often inconsistent with those deduced by experts through their domain knowledge, which can be confusing to those experts. This work introduces RASIPAM, a \textbf{RA}cket-\textbf{S}ports \textbf{I}nteractive \textbf{PA}ttern \textbf{M}ining system, which allows experts to incorporate their knowledge into data mining algorithms to discover meaningful tactics interactively. RASIPAM consists of a constraint-based pattern mining algorithm that responds to the analysis demands of experts: Experts provide suggestions for finding tactics in intuitive written language, and these suggestions are translated into constraints to run the algorithm. RASIPAM further introduces a tailored visual interface that allows experts to compare the new tactics with the original ones and decide whether to apply a given adjustment. This interactive workflow iteratively progresses until experts are satisfied with all tactics. We conduct a quantitative experiment to show that our algorithm supports real-time interaction. Two case studies in tennis and in badminton respectively, each involving two domain experts, are conducted to show the effectiveness and usefulness of \mbox{the system.}
} 


\begin{document}



\begin{spacing}{0.98}
\firstsection{Introduction}

\maketitle

\begin{figure}[tb]
  \centering 
  \includegraphics[width=\columnwidth]{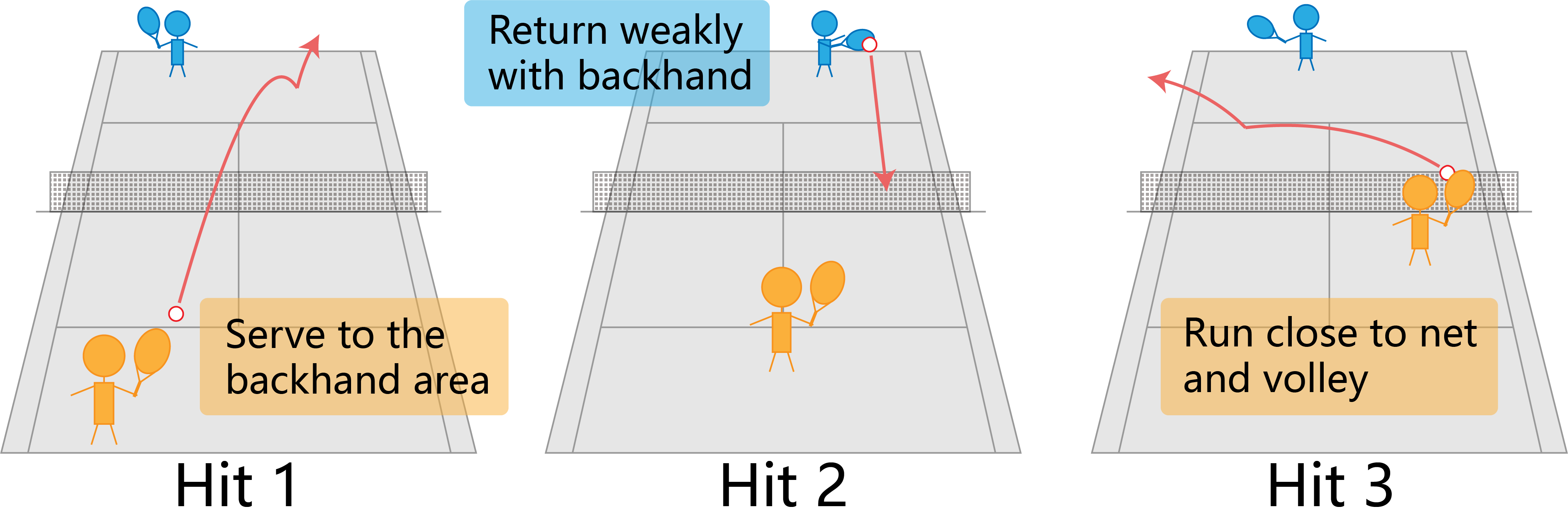}
  \vspace{-0.25in}
  \caption{An example of a tennis tactic with three hits --- ``serve-and-volley.''}
  \vspace{-0.15in}
  \label{fig:tactic_example}
 \end{figure}

Coaches and data analysts working in racket sports like tennis and badminton use tactical analysis to gain high-level insight into these sports, such as players' preferred playing styles and competitive weaknesses \cite{wang2021tac}.
In racket sports, a \textit{rally} starts with one player serving the ball, progresses as two players \textit{hit} the ball in alternation, and ends with one player scoring a point.
For each hit, a player must consider multiple hit features such as the hitting technique and the ball position.
By skillfully combining different hit features over several consecutive hits, a player may adopt a \textit{tactic} for winning the game.
Fig. \ref{fig:tactic_example} demonstrates an example in tennis -- Player 1 (P1) may play a three-hit ``serve-and-volley'' tactic against Player 2 (P2).
\textit{Hit 1}: P1 serves the ball to the backhand area \textit{(ball position)} of P2;
\textit{Hit 2}: P2 hits a weak return via his/her backhand \textit{(hitting pose predicted by P1)};
\textit{Hit 3}: P1 runs close to the net \textit{(player position)}, volleys the ball back \textit{(hitting technique)}, and win a point.
To help experts quickly find such tactics, some data-driven methods \cite{wu2021tacticflow,wang2021tac} model each hit as a multivariate event and each rally as an event sequence.
By applying pattern mining algorithms, such methods can discover recurring patterns and analyze them as tactics.

However, domain experts' knowledge can lead them to interpret the results of data-driven methods in specific ways.
For instance, if an algorithm finds that P1 frequently performs the aforementioned ``serve-and-volley'' tactic, experts may interpret this finding differently based on what they know about P1.
If P1 prefers net tactics, experts may want to focus on the third hit, where P1 runs close to the net to volley the ball.
Otherwise, experts may \revise{R4C8.A}{focus on the first hit}{want to merge tactics with a similar first hit}, where P1 hits the ball to P2's backhand, preventing P2 from returning the ball effectively.
Experts use the domain knowledge gathered over the years to perform such fine-grained analysis.
\revise{R4C8.A}{}{Lacking this domain knowledge, data-driven algorithms cannot support such further adjustments, such as merging tactics with specific similar hits. Thus, the mined tactics are inconsistent with what experts know, confusing and difficult to use\cite{liu2021mtv,wang2021bridging}.}
\revise{SRC4,R2C4}{}{An experiment presented in Sect. \ref{sec:evaluation} also suggested that few tactics expected by experts could be directly and precisely mined by data-driven algorithms.}
Experts would prefer to work interactively with algorithms, giving their knowledge-based suggestions so that the algorithm's findings better meet their needs.

However, although interactive pattern mining is well-studied \cite{boley2013one,liu2019tpflow,ming2019protosteer,andrienko2021theoretical}, it is still challenging to apply it to multivariate tactics.
One issue is \textbf{how to handle experts' fine-grained suggestions on multivariate tactics.}
Experts may give meticulous feedback about the algorithm's treatment of a single hit feature because small details may determine the outcome of a highly competitive game.
To the best of our knowledge, no work has studied interactive pattern mining that includes such fine-grained adjustments.
\revise{R4C2}{}{Another issue is \textbf{how to ensure experts can intuitively evaluate the adjusted tactics.}
After the algorithm adjusts the tactics, experts need to explore the adjusted tactics and judge whether the adjustment is reasonable.}
However, because multivariate tactics \revise{R3C8.B}{are so complex}{involve so many hit features of many hits}, understanding an adjustment that may involve changes among many tactics (e.g., merging several similar tactics) is extremely difficult and time-consuming.
Intuitive methods are required to help experts quickly verify the results.

To solve these issues, we propose RASIPAM, a \textbf{RA}cket-\textbf{S}ports \textbf{I}nteractive \textbf{PA}ttern \textbf{M}ining system.
To tackle the first challenge, we \textbf{propose a constraint-based pattern mining algorithm}.
We conduct a pre-study to collect experts' suggestions and delimit the space where suggestions occur.
We further define a set of natural language templates to transfer experts' suggestions into math representations, i.e., constraints.
The algorithm can generate candidate patterns based on these constraints and then use a data-driven metric to select the best candidates for experts, combining domain knowledge with a data-driven method.
To address the second challenge, we \textbf{propose an interactive visual interface} with a comparative visual design.
A semantic-based projection view and a glyph design for visualizing multivariate hit are proposed to allow experts to evaluate the adjustment by comparing the new tactics with the original ones, with multiple levels of detail \cite{shneiderman1996eyes}.

\revise{SRC4,R2C4}{}{
Finally, we present two quantitative experiments evaluating the quality of the adjusted tactics and the speed of the algorithm, which demonstrate our algorithm's effectiveness, efficiency, and necessity.
We also evaluate the usability of our system through two case studies, performed with two experts each from tennis and badminton, which prove that our system can effectively incorporate expert knowledge to interactively mine meaningful tactics for domain experts.
}

In summary, this work has the following contributions:
\revise{SRC6,R3C1}{}{
\begin{itemize}[nosep, leftmargin=*]
    \item A design study in close collaboration with experts in racket sports, which helps to carve the domain requirements on interactive tactic mining and summarize nine potential adjustments to tactics.
    \item A human steerable multivariate pattern mining algorithm that allows fine-grained adjustment of patterns based on experts’ suggestions.
    \item A user interface that contains a natural language interface to help experts quickly adjust tactics and a comparative visual design to help experts evaluate the adjusted tactics.
    \item Two quantitative experiments that evaluate the effectiveness and efficiency of the algorithm, and two case studies on tennis and badminton that demonstrate our system’s usefulness.
\end{itemize}
}

\section{Related Work}


\subsection{Interactive Pattern Mining for Event Sequences}

Methods based on Sequential Pattern Mining (SPM) find all subsequences whose frequency is higher than a threshold \cite{fournier2017survey}.
Due to the well-known problem of \textit{pattern explosion} \cite{menger2015experimental}, users usually need to adjust the threshold manually and constantly to obtain a suitable number of patterns, where interactive mining algorithms are needed \cite{aggarwal1998online}.
Many SPM-based algorithms \cite{parthasarathy1999incremental,lin2003improving,ren2006fast} cached mined patterns while incrementally mining new patterns.
Wojciechowski et al. \cite{wojciechowski2001interactive} extended the adjustment to \revise{R4C8.D}{the size and length of patterns}{the number of values/events in each pattern}.
However, SPM-based methods may generate similar and repetitive patterns that are frequent but meaningless for tactical analysis \cite{menger2015experimental}.

Interactive machine learning (IML) has been much-discussed recently because it incorporates user feedback into model computations, enhancing the interpretability of the results \cite{dudley2018review}.
ProtoSteer \cite{ming2019protosteer} proposed a tailored interactive pattern mining system based on a deep sequence model \cite{ming2019interpretable}.
However, training an effective model for mining racket sports tactics is difficult, because players have unique tactics and continually change them as needed to win against particular opponents.

Over the past two decades, algorithms based on the Minimum Description Length (MDL) principle have emerged \cite{grunwald2007minimum}.
Unlike SPM-based methods, MDL-based methods find a small set of optimized patterns instead of surfacing all the patterns that appear frequently.
In addition, unlike IML-based methods, MDL-based methods do not rely on training data.
Meanwhile, many MDL-based methods have investigated how to incorporate domain knowledge into pattern mining.
Tatti et al. \cite{tatti2010using} allowed users to define constraints to support interactive analysis.
One-Click Mining \cite{boley2013one} could generate new patterns based on users' binary feedback (like/dislike) about current patterns.
However, to the best of our knowledge, no MDL-based methods have studied interactive mining of multivariate event sequences.
Our work proposes an MDL-based algorithm for mining multivariate patterns interactively.

\subsection{Visual Analytics of Multivariate Event Sequences}

A comprehensive survey by Guo et al. \cite{guo2021survey} summarizes a wealth of research on visual analytics of event sequences, which considers multivariate event sequence visualization as a challenging and promising research direction.
Some visualizations directly displayed high-dimensional data in detail as matrices \cite{loorak2015timespan} and lines \cite{kim2021skyflow}.
PatternFinder \cite{fails2006visual} and EventPad \cite{cappers2017exploring,cappers2018eventpad} propose novel interfaces to query multivariate sequential patterns.
Wang et al. \cite{wang2015visual,wang2017visual} applied causality analysis to multivariate event sequence data.
Many researchers have applied visual analytics to sports data because it is a common type of multivariate event sequence data.
Wu et al. \cite{wu2017ittvis} used a sequence of matrices to visualize intra-event relationships (i.e., among multiple attributes) as well as inter-event ones.
\revise{SRC1,R2C1,R1C1}{}{Tac-Miner \cite{wang2021tac} and Tac-Simur \cite{wang2019tac} designed intuitive glyphs to encode multivariate events in table tennis, which inspired our work to design glyphs tailored for tennis and badminton. Based on the glyph design, our work mainly contributes to a comparative visual design for comparing the adjusted tactics with the original ones, allowing experts to evaluate their suggested adjustments.}

\subsection{Visual Analytics of Sports Tactics}

Tactics are sets of high-level decisions made by players, which can significantly affect the outcome of games \cite{hibbs2013strategy}, leading to experts' needs for visual analytics \cite{du2021survey,wu2022defence,stoiber2022perspectives}.
In team sports like soccer, tactics are mainly related to the team members' cooperation\cite{cao2021mig}, such as the formation changes \cite{wu2018forvizor} and multiple players' movement coordination \cite{andrienko2019constructing,xie2020passvizor}.
In racket sports, tactical analysis mainly focuses on correlations among consecutive hits\cite{Lan2021RallyComparator,Chu2021TIVEE}.
\revise{SRC1,R2C1,R1C1}{}{Wang et al. \cite{wang2021tac} and Wu et al. \cite{wu2020visual,wu2021tacticflow} proposed data-driven algorithms for mining multivariate tactics. These algorithms analyze tactics in a data-driven way by proposing novel models to discover insights automatically for users. Our work further incorporates experts’ knowledge into the algorithm to mine meaningful tactics. We mainly contribute to a pre-study that collects hundreds of experts’ potential suggestions on tactics and a human steerable tactic mining algorithm that can fine-tune tactics based on these suggestions.}

\section{Design Study}

This section presents a design study for summarizing the analysis tasks involved with interactive tactical analysis, and proposes the analysis workflow and system architecture for performing these tasks.

\subsection{Setup and Study Process}

Our study follows the design study methodology of Sedlmair et al. \cite{seldmair2012design}.

\textbf{Experts.}
We collaborated with five domain experts in three racket sports --- tennis (E1, E2), badminton (E3, E4), and table tennis (E1, E5) --- \textit{E1 focuses on table tennis but is also proficient in tennis}.
All experts were professional athletes and are now sports science data analysts with more than five years of experience.

\textbf{Process.}
Over the past year, we held weekly meetings with these five experts to discuss interactive tactical analysis in three steps.

\textit{Step 1:}
Given that each sport has domain-specific analysis methods, we interviewed experts separately according to their focal sports.
For each sport, we first confirmed the need for interactive tactical analysis.
We then gathered each expert's concerns about and suggestions for interactive tactical analysis.
Each interview lasted about one hour.

\textit{Step 2:}
By summarizing the experts' feedback, we further formalized the analysis tasks for all racket sports (\ref{task:overview}-\ref{task:sequence} in Sect. \ref{sec:tasks}), which were further discussed and improved in three meetings with all the experts.
Due to commonalities between all racket sports, most experts' concerns could be covered by the same task across the different sports (\ref{task:overview}-\ref{task:similarity}).
However, experts from different sports cared about different details.
For example, E5 preferred to know the frequency of tactics at the hit-by-hit level because rallies in table tennis are usually short, meaning every hit matters.
But E4 did not care about this because of the long rallies in badminton.
We covered all these detailed concerns with \ref{task:sequence}.

\textit{Step 3:}
We developed a prototype system for domain experts, along with real use cases, to further gather their feedback on the analysis tasks and the system design.
After two months of polishing, we finalized our analysis tasks and obtained the current version of the system.

\subsection{Task Analysis}

\label{sec:tasks}

The main tasks our system needs to perform during interactive tactical analysis are as follows, where \ref{task:overview}-\ref{task:similarity} are common to all racket sports.

\begin{enumerate}[nosep,label={\bf T{{\arabic*}}},leftmargin=*]
    \item \label{task:overview}
    \textbf{Get an overview of an individual player's tactics.}
    Experts prefer player-centric tactical analysis, because each player uses unique tactics.
    Rather than analyzing each tactic of a player directly, experts expect to start with an overview to learn things such as how many tactics result in more wins, and whether the player specializes in a few tactics or adopts more tactics to confuse opponents.
    \item \label{task:adjust}
    \textbf{Adjust mined tactics according to experts' suggestions.}
    Experts expect to give knowledge-based suggestions to the algorithm about how to better mine tactics.
    For example, the ``seesaw battle'' tactic (i.e., \textit{both players hitting the ball with a drive technique}) is common in tennis games, and is easily detected by the algorithm.
    However, experts may argue that the ball position determines the outcome of a seesaw battle and require our algorithm to subdivide this tactic into several subtactics based on different ball position changes.
    Because of the low frequency of each subdivided tactic, a data-driven algorithm usually fails to directly discover them.
    \item \label{task:evaluation}
    \textbf{Evaluate the adjusted results.}
    Experts' knowledge comes from past games, and may lose usefulness as a player's tactics evolve over time.
    Just as the experts suggest adjustments, experts expect our system to evaluate their suggestions --- \textit{providing data-driven metrics} and \textit{comparing adjusted tactics with the original ones} --- to prevent them from making inappropriate adjustments.
    For example, when experts find a tactic not typical of a certain player and suggest removing it, our system could provide data indicating that this may be a newly developed tactic rather than anomaly, thus reminding experts to exercise caution before discounting it.
    \item \label{task:similarity}
    \textbf{Discover similar tactics.}
    Players often achieve different tactical goals (e.g., confusing their opponents) by changing a few hit features of a previous tactic, resulting in a slightly different tactic. Finding these tactics is valuable for tactical analysis because it helps experts study which changes are effective. Meanwhile, merging similar tactics that actually lead to similar outcomes can help experts avoid spending their analysis time unnecessarily.
    \item \label{task:sequence}
    \textbf{Display raw sequences in detail.}
    The raw sequences can also include details about the context of a tactic, i.e. the previous and subsequent hits, which indicate the impetus for and the results of using the tactic, respectively.
    Experts expressed a desire to explore the raw sequences by viewing statistics (E1, E5), observing the detailed hit features of each hit (E1 - E5), and watching videos (E1 - E4), in order to obtain a deep understanding of the tactic.
\end{enumerate}

\begin{figure}[tb]
    \centering 
    \includegraphics[width=\columnwidth]{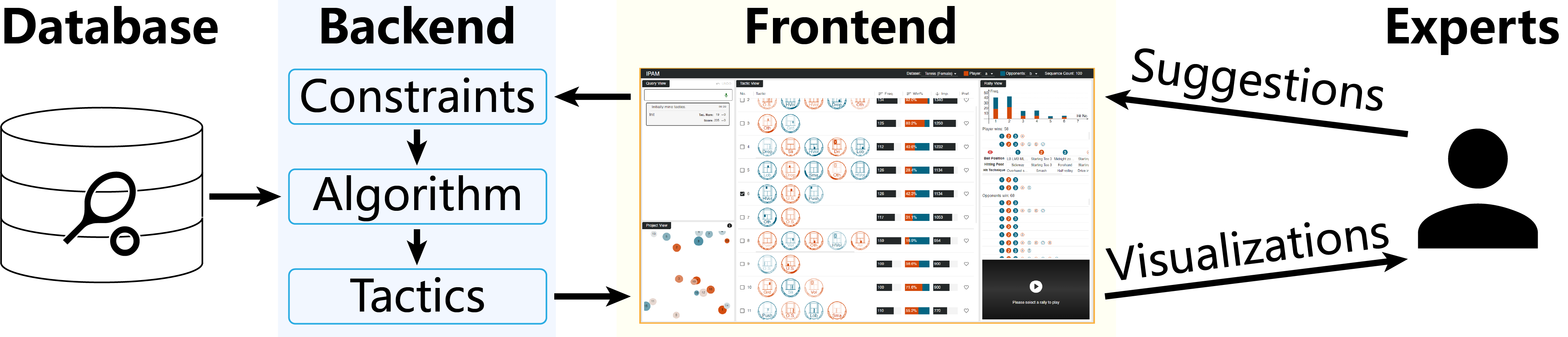}
    \vspace{-0.2in}
    \caption{Our system architecture for interactive tactical analysis.}
    \vspace{-0.15in}
    \label{fig:architecture}
   \end{figure}

\subsection{Analysis Workflow}

\begin{figure*}[!htb]
    \centering 
    \includegraphics[width=\linewidth]{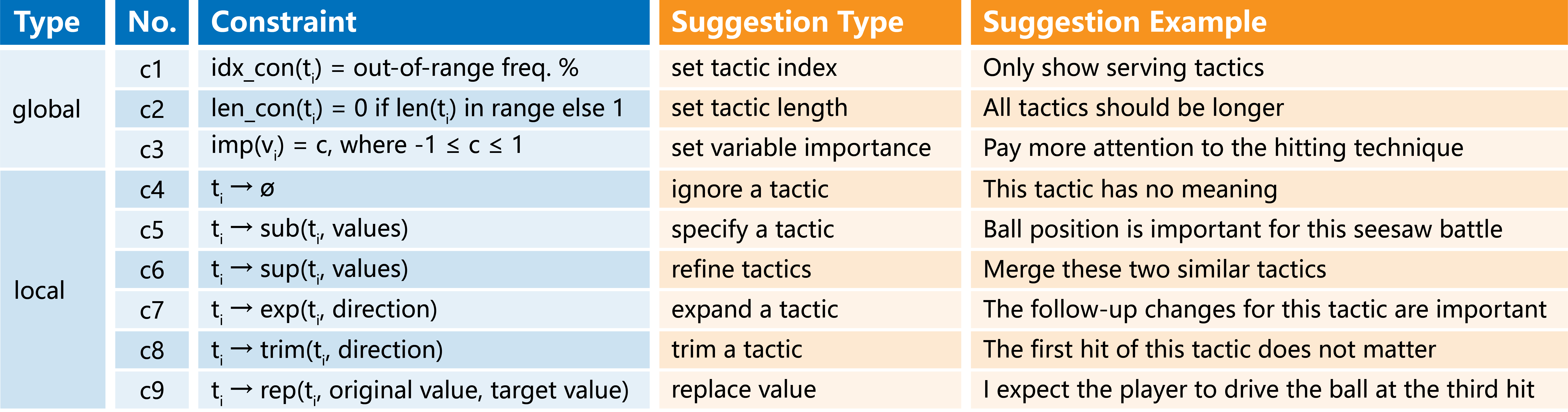}
    \vspace{-0.2in}
    \caption{The constraint space with nine constraints, each of which supports one type of suggestion, demonstrated through an example given by experts. \revise{R3C3}{}{Three global constraints satisfy experts' limitations for the overall tactic set, while six local ones help experts fine-tune specific tactics.}}
    \vspace{-0.15in}
    \label{fig:constraints}
   \end{figure*}


We propose an analysis workflow based on a human-in-the-loop architecture (Fig. \ref{fig:architecture}) to address the tasks delineated above, which are implemented through an open-source prototype system\footnote{\href{https://anonymous.4open.science/r/Rasipam-A695/}{https://anonymous.4open.science/r/Rasipam-A695/}. Only synthetic data exists due to the confidentiality of real-world datasets.} consisting of a database, a backend, and a frontend.
The \textbf{database} stores the raw sequences from different racket sports (Sect. \ref{sec:data_structure}).
The \textbf{backend} runs a mining algorithm (Sect. \ref{sec:algorithm}) to discover a set of tactics that satisfy the constraints suggested by experts (Sect. \ref{sec:constraints}).
The \textbf{frontend} visualizes the tactics for experts and collects their knowledge-based suggestions in return (Sect. \ref{sec:interface}).
Our workflow starts when the backend mines an initial tactic set from the dataset, which are then visualized by the frontend (\ref{task:overview}, \ref{task:similarity}, \ref{task:sequence}).
Next, experts explore the visualizations and suggest our algorithm to adjust these tactics (\ref{task:adjust}).
Our system provides evaluation of the adjusted results (\ref{task:evaluation}), helping experts preview the results and determine whether to apply the adjustment.
Experts can give suggestions iteratively until they have no more \revise{R3C8}{concerns}{suggestions}.

\section{Data Model}

In this section, we introduce the data structures of the raw sequences and tactics. We then delimit the space of domain-specific constraints, and describe the pre-study we performed to obtain those constraints.

\subsection{Data Structure}

\begin{figure}[tb]
    \centering 
    \includegraphics[width=\columnwidth]{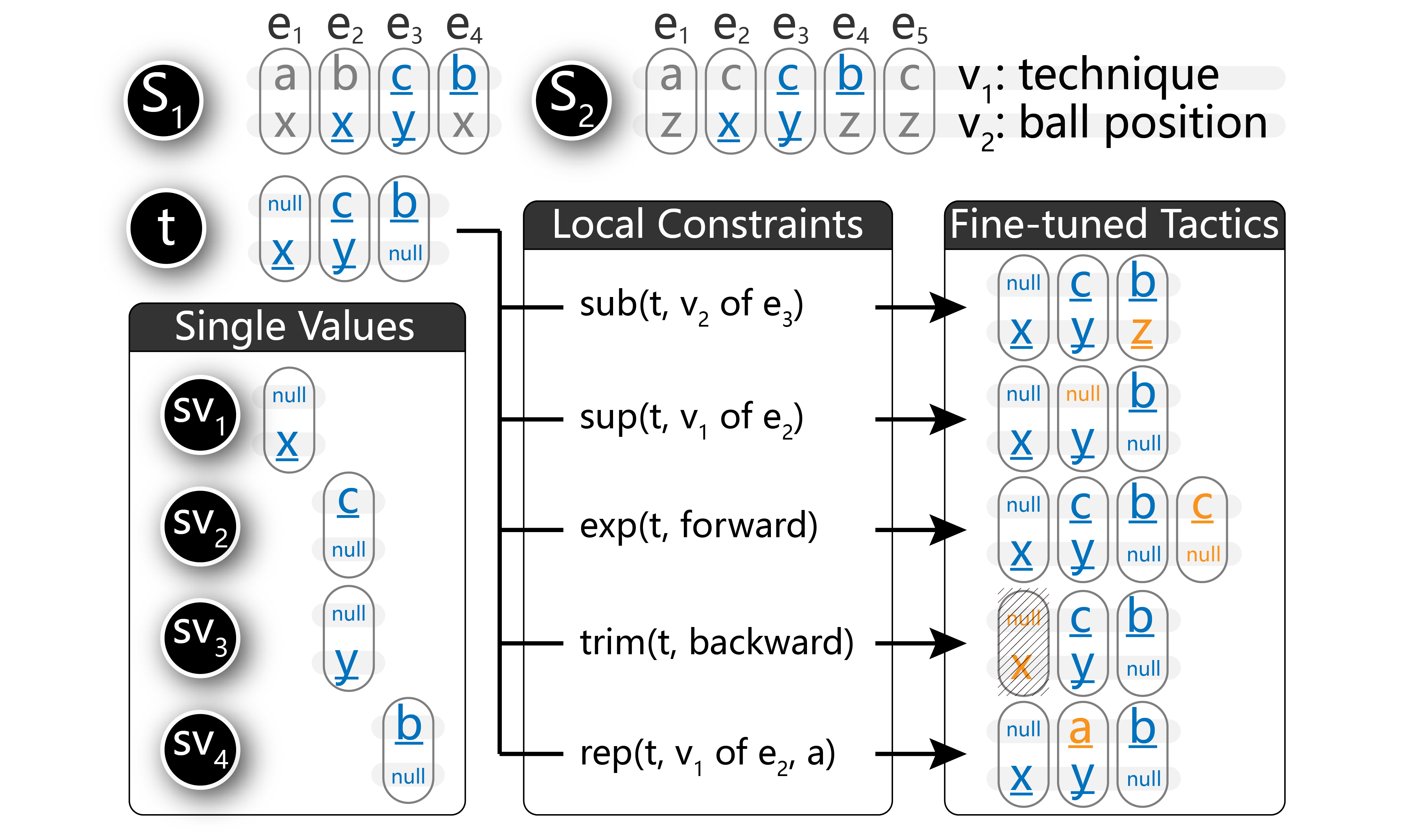}
    \vspace{-0.2in}
    \caption{Sequences $s_1$ and $s_2$ both involve two hit features, where tactic $t$ is used (highlighted in blue). We further present how each local constraint affects the tactic (highlighted in orange).}
    \vspace{-0.15in}
    \label{fig:tactic_definition}
   \end{figure}

\label{sec:data_structure}

\textbf{Raw Sequences:}
A dataset consists of $n$ raw sequences, denoted $S=\{s_1, s_2, ..., s_n\}$.
Each sequence $s=\{e_1, e_2, ..., e_l\}$ represents a rally with $l$ hits, \revise{R4C6}{}{with the player who served the ball and the player who won the point as metadata.}
Each event $e=\{v_1, v_2, ..., v_k\}$ represents a hit with $k$ categorical values indicating multiple hit features.

\textbf{Tactics:}
At the beginning of the process, and again after each adjustment, our algorithm outputs a tactic set $T=\{t_1, t_2, ..., t_m\}$ with $m$ tactics.
Each tactic $t_i$ is defined as a multivariate subsequence shared by many rallies, which is \textit{value-nullable} and \textit{consecutive}.
\revise{R3C2,R4C7.B}{}{
For example, in Fig. \ref{fig:tactic_definition}, tactic $t$ is applied in both $s_1$ and $s_2$ (highlighted by blue).
Tactic $t$ consists of four \textit{single values} from $sv_1$ to $sv_4$ (each of which is a non-null value) and two null values (meaning that the corresponding values are different in $s_1$ and $s_2$).
Meanwhile, in $s_1$ and $s_2$, $t$ is always used consecutively without any hits in between.
}
\revise{R4C6}{}{In addition, each tactic records the reference to the rallies and the hits where the tactic was applied as the metadata.}

\subsection{Constraints}

\label{sec:constraints}

Many previous works have studied generic constraints for mining univariate patterns \cite{wojciechowski2001interactive,tatti2010using}.
However, for the multivariate tactics in racket sports, we need fine-grained and domain-specific constraints to satisfy experts' suggestions.
Thus, we conducted a pre-study with the five experts (E1-E5) to find all the constraints needed, i.e., the constraint space.
The process of this study and its results are as follows.

\subsubsection{Study Process}

\label{sec:prestudy}

The study was divided into three stages.

\textbf{Preparation Stage:}
We prepared a real tactic set for the experts to spur the suggestions that may arise in real analysis scenarios, which contains 107/102/95 tactics mined from 3,052/3,123/2,997 rallies played between top players in tennis/badminton/table tennis respectively, by an open-source tactic mining algorithm\cite{wu2021tacticflow}.
We also prepared an initial constraint space from the perspective of data mining, with several example suggestions for each constraint for experts reference.

\textbf{Collection Phase:}
The five experts were invited to give suggestions on our prepared tactics -- each tactic was reviewed by at least one expert.
We first showed them the example suggestions and encouraged them to give suggestions beyond the examples, as long as the suggestions fit their actual needs.
Their suggestions and the tactics they expected were recorded for further analysis.

\textbf{Summarization Phase:}
We filtered 297 different suggestions from the 349 ones collected and classified them into nine suggestion types (Fig. \ref{fig:constraints}).
We further expanded the constraint space to cover all these suggestions and held discussions with experts to ensure comprehensive support of possible suggestions from experts.

\begin{figure*}[tb]
    \centering 
    \includegraphics[width=\linewidth]{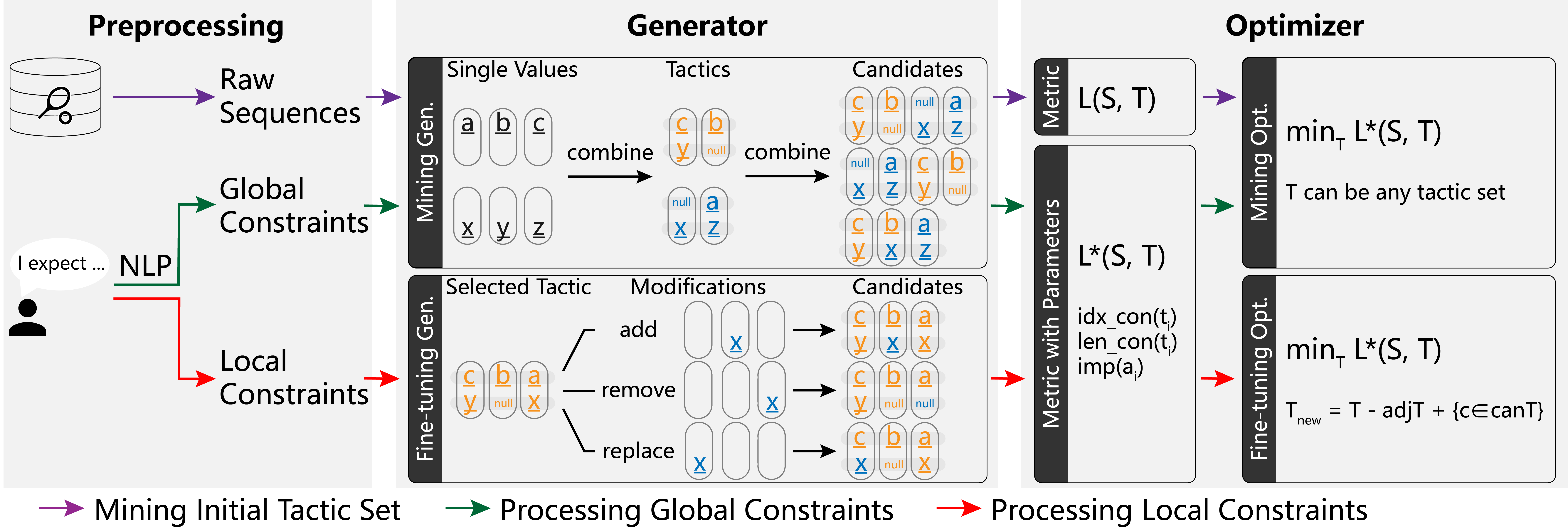}
    \vspace{-0.2in}
    \caption{Algorithm framework for mining tactics in three scenarios. \purpleArrow: the algorithm can mine an initial tactic set from raw sequences. \greenArrow: the algorithm proposes several parameters to process global constraints and re-mine the tactic set. \redArrow: the algorithm proposes a fine-tuning generator and a fine-tuning optimizer to fine-tune specific tactics based on local constraints.}
    \vspace{-0.15in}
    \label{fig:algorithm}
\end{figure*}

\subsubsection{Results}

We list the final constraints and the supported suggestion types in Fig. \ref{fig:constraints} and elaborate how we apply each constraint in Sect. \ref{sec:alg_c}.
Due to space limitations, we will give a brief introduction here and present more details in the appendix.
We classify the constraints into two types according to their effects.

\textbf{Global constraints} affect the whole tactic set, leading to an adjusted tactic set that is completely different from the original one (thus, \ref{task:evaluation} is not applicable).
\revise{R3C3}{}{
Taking c1 in Fig. \ref{fig:constraints} as an example, experts may only want to analyze serving tactics (\textit{the expert's suggestion}).
Rather than just filtering a few serving tactics from the original tactic set, the system should mine a new set of serving tactics for detailed analysis (\textit{limiting the index of the tactics' first hit to 1}).
To mine only serving tactics, our algorithm will adds a loss term $idx\_con(t_i)$ to an objective function, giving a high cost to other tactics (\textit{the constraint, which is elaborated in Sect. \ref{sec:alg_c}}).
}

\revise{R3C3}{}{
\textbf{Local constraints} require fine-tuning of a few tactics, without changing other tactics.
Taking c6 in Fig. \ref{fig:constraints} as an example, experts may suggest merging two similar tactics.
The system should then refine the original two tactics to a super-tactic that contains all common values in the tactics to be merged.
In this case, our algorithm will add a constraint ``$t_i \rightarrow sup(t_i, values)$'' to each original tactic, which means replacing the original tactic $t_i$ with the super-tactic $sup(t_i, values)$, where $values$ represent the common values in the two original tactics.
}
\section{Interactive Tactic Mining Algorithm}

\label{sec:algorithm}

We propose an interactive multivariate pattern mining algorithm to meet experts' tactic-mining needs in three scenarios as follows (Fig. \ref{fig:algorithm}).
\revise{SRC3,R2C3}{}{
\begin{enumerate}[nosep,label={\bf S{{\arabic*}}},leftmargin=*]
    \item \label{sce:initial} Initially, the algorithm needs to mine an initial tactic set $T$ from the rallies dataset $S$, saving experts’ time needed to discover tactics manually (indicated by \purplearrow).
    Given the dataset $S$ and the current tactic set $T$, experts may iteratively suggest constraints to adjust $T$, leading to the next two scenarios.
    \item \label{sce:global} To satisfy the global constraints, our algorithm needs to re-mine a new tactic set $T_{new}$ from dataset $S$ (indicated by \greenarrow).
    \item \label{sce:local} To satisfy experts' local constraints, our algorithm needs to fine-tune several specific tactics in $T$, while leaving other tactics unchanged (indicated by \redarrow).
\end{enumerate}
To accomplish this, we extend the existing algorithm of TacticFlow \cite{wu2021tacticflow}, which can mine an initial tactic set (for \ref{sce:initial}). The main contributions of our algorithm lie in processing global and local constraints to enable interactive tactic mining (for \ref{sce:global} and \ref{sce:local}).}

\subsection{Initializing Tactic Set}

\revise{SRC3,R2C3}{}{The algorithm receives a raw sequence dataset $S$ as input and tries to mine a set of initial tactics $T$ from $S$ as output.}
\revise{R4C7.B}{}{The mining process is based on the MDL principle, which regards $T$ as a model to describe $S$ and regards the model that results in the minimum description length (i.e., a metric for computing the information cost \cite{grunwald2007minimum}) as the best option.}
The algorithm consists of a generator and an optimizer.
The generator randomly combines two tactics in the tactic set at different alignments to generate new candidates, starting from combining single values, such as the \textit{drive} technique, the \textit{backcourt} ball position, and so on.
Every time the generator generates a candidate, the optimizer uses a domain-specific metric to compute the description length $L(S, T)$ (Eq. \ref{eq:oriDL}), which mainly considers the number of tactics $|T|$, the frequency of each tactic $freq(t_i)$, and the single values $sv(s_i)$ that cannot be described by the tactics in each sequence.
\revise{SRC3,R3C4}{}{Constants $\alpha$ and $\beta$ control the weight of each term to ensure that each term matters, considering that experts generally prefer to analyze about 20 tactics (making $|T|$ small), which are used many times to describe hundreds of sequences (making $freq(t_i)$ large) and leave thousands of single values (making $sv(s_i)$ very large).}
The optimizer further optimizes the tactic set by adding candidates that can reduce the description length into the tactic set and removing tactics that cannot benefit from optimizing the description length from the tactic set.
The generator and optimizer work iteratively until the tactic set does not change in a certain iteration, whereupon the algorithm obtains a tactic set with the appropriate minimum description length.

\begin{equation}
    \begin{aligned}
        L(S, T) = |T| + \alpha \sum_{t_i \in T}freq(t_i) + \beta \sum_{s_i \in S}sv(s_i),
    \end{aligned}
    \label{eq:oriDL}
\end{equation}

\subsection{Processing Constraints}

\label{sec:alg_c}

Our algorithm contributes to two workflows --- one each for processing global and local constraints.
Global constraints require changing the entire tactic set, re-defining what kind of tactic set is the best (i.e., the one expected by experts).
Following the MDL principle, which associates the best model with the minimum description length, we change the metric (``Metric with Parameters'' in Fig. \ref{fig:algorithm}) to discover the new tactic set regarded as the best.
In contrast, local constraints require the algorithm to fine-tune specific tactics, leaving others unchanged.
We propose a new generator and optimizer for fine-tuning (``Fine-tuning Gen.'' and ``Fine-tuning Opt.'' in Fig. \ref{fig:algorithm}).

\subsubsection{Global Constraints}

We introduce a new metric $L^*(S, T)$ (Eq. \ref{eq:newDL}) to fit the global constraints.
\revise{R3C3}{}{The new metric adds three loss terms, giving a high cost to the tactics that do not fit the global constraints so that the optimizer prefers to remove them.}
This mainly involves three parameters as follows.

\begin{itemize}[nosep, leftmargin=*]
    \item \textbf{c1}: $idx\_con(t_i)$.
    For each tactic $t_i$ in $T$, we find the usages at in-range indexes (e.g., 1-4 for serving tactics) and out-of-range ones and calculate the percentage of out-of-range usages as $idx\_con(t_i)$.
    We multiply the frequency of each tactic with $idx\_con(t_i)$ to give a high cost for out-of-range usages.
    \item \textbf{c2}: $len\_con(t_i)$.
    For each tactic $t_i$, $len\_con(t_i)$ represents whether the length of $t_i$ occurs in a range expected by experts (e.g., $>3$ for tactics expected to include more than three hits) --- 0 means in range, and 1 means out of range. We multiply the frequency of each tactic with $len\_con(t_i)$ to give a high cost for tactics with out-of-range length.
    \item \textbf{\revise{SRC3,R3C4}{c3 and c4}{c3}}: $imp(v_i)$.
    We denote the importance of each hit feature value as $imp(v_i)$, ranging from -1 to 1 and defaulting to 0. We apply \revise{SRC3,R3C4}{c3 and c4}{c3} by \revise{SRC3,R3C4}{timing}{multiplying} the number of single values with the importance of the corresponding hit feature, giving a high cost to any single values that cannot described by tactics but are regarded important by experts.
\end{itemize}
\revise{SRC3,R3C4}{}{To make our metric L*(S, T) comparable with L(S, T), we directly use the same parameters alpha and beta as TacticFlow to control the weight of each term.}
Note that, initially with no global constraints, $idx\_con(t_i) = len\_con(t_i) = imp(v_i) = 0$, leading to $L^*(S, T) = L(S, T)$.

\begin{equation}
    \begin{aligned}
        &L^*(S, T) = L(S, T) + \alpha \sum_{t_i \in T}freq(t_i) \times idx\_con(t_i)\\& + \alpha \sum_{t_i \in T}freq(t_i) \times len\_con(t_i)) + \beta \sum_{s_i \in S}\sum_{v_i}sv(v_i, s_i) \times imp(v_i),
    \end{aligned}
    \label{eq:newDL}
\end{equation}

\subsubsection{Local Constraints}

We propose a modification-minimized generator to generate new candidates that satisfy the local constraints based on a tactic $t$.
Meanwhile, we propose a local sensitive optimizer to fine-tune certain aspects of the tactic set while leaving other tactics unchanged.

\textbf{Generator.}
\revise{SRC3,R3C4}{}{
Based on the tactic to adjust, the generator uses a BFS algorithm to search for all candidates that satisfy the local constraints (Fig. \ref{fig:tactic_definition}), applying a minimum number of local modifications.
Each modification (shown as ``Fine-tuning Gen.'' in Fig. \ref{fig:algorithm}) consists of a target value to change and an action selected from \textit{add}, \textit{remove}, and \textit{replace}.
For example, to trim the last hit with two non-null values from a tactic (c8 in Fig. \ref{fig:constraints}), the generator will generate one candidate by applying two modifications -- removing the first value and the second value in the last hit, respectively.
As another example, to expand a tactic with one more hit (c7 in Fig. \ref{fig:constraints}), the generator will apply only one modification -- adding one value to the back of the tactic.
But there usually exist numerous values that can be added, which lead to many candidate tactics.
}
In this way, the generator accurately generates candidates that meet experts' expectations with the minimum modifications, making it easy for experts to compare the new tactics with the original ones.

\textbf{Optimizer.}
\revise{SRC3,R3C4}{}{
The fine-tuning optimizer is local sensitive, which only replaces the tactics suggested to adjust with the generated candidates that can minimize the description length, keeping other tactics unchanged (``Fine-tuning Opt.'' in Fig. \ref{fig:algorithm}).}
Formally, the optimizer tries to find the best tactic set \revise{SRC3,R3C4}{}{$T_{new} = T - adjT + \{c \in canT\}$}, where $T$ is the tactic set to adjust, $adjT$ is the tactics experts would like to adjust, and $canT$ is all the candidates generated by the generator.
The optimizer performs in three steps as follows (Alg. \ref{alg:optimizer}).
\begin{enumerate}[nosep,leftmargin=*]
    \item \textit{Line 1:}
    \revise{SRC3,R3C4}{}{The algorithm removes all tactics that experts have suggested adjusting from the original tactic set.}
    \item \textit{Line 2-4:}
    \revise{SRC3,R3C4}{}{The algorithm iterates all the candidates, with high-frequency ones first.
    If a candidate facilitates reducing the description length, it will be added to the tactic set.
    The optimizer keeps the same metric $L^*(S, T)$ used when applying global constraints so that the description length is comparable.}
    \item \textit{Line 5-7:}
    \revise{SRC3,R3C4}{}{Every time the algorithm adds a candidate, it prunes the tactic set by removing redundant candidates (i.e., those candidates whose inclusion leads to a larger description length), leaving other tactics not related to the constraints unchanged.}
\end{enumerate}
There exists a trade-off for interactive analysis --- we ensure that at least one candidate will be added to the tactic set (line 3 and line 5), satisfying experts' needs for adjustments, while the description length may be larger than the original tactic set.
A small increase is acceptable for the trade-off, but a large one may indicate that experts have made an inappropriate adjustment, which should be prevented.
Thus, we provide experts with data-driven metrics for scoring a tactic set and evaluating the tactical importance of each tactic.
\revise{SRC3,R3C4}{}{We score a tactic set by $score(S, T) = L(S, \emptyset) - L(S, T)$ to provide experts with a more intuitive evaluation on a tactic set than $L(S, T)$ because $score(S, T)$ is usually much smaller than $L(S, T)$ (hence easy to read and compare) and is positively correlated to the quality of the tactic set (i.e., a higher score means a better tactic set).}
Similarly, we evaluate the tactical importance of a tactic by $tac\_imp(t) = L(S, T) - L(S, T - t)$, where tactics with higher tactical importance are more valuable.

\begin{algorithm}[b]
    \label{alg:optimizer}
    \caption{Fine-tuning Optimizer}
    \DontPrintSemicolon

    \KwInput{Sequence set $S$, current tactic set $T$,\linebreak tactics to adjust $adjT$, candidate tactics $canT$}
    \KwOutput{New tactic set $T_{new}$}

    $T_{new} = T - adjT$\;
    \For{$ct \gets canT$ in order of frequency}
    {
        \If{$L^*(S, T_{new} + ct) < L^*(S, T_{new})$ \textbf{or} $canT \bigcap T_{new} \neq \emptyset$}
        {
            $T_{new} = T_{new} + ct$\;
            \For{$t \gets canT \bigcap T_{new}$ \textbf{and} $t \neq ct$}
            {
                \If{$L^*(S, T_{new} - t) \le L^*(S, T_{new})$}
                {
                    $T_{new} = T_{new} - t$\;
                }
            }
        }
    }
    \Return $T_{new}$
\end{algorithm}
\section{Visual Interface}

\label{sec:interface}

Our proposed visual interface consists of a control bar (Fig. \ref{fig:teaser}A) and four main views (Fig. \ref{fig:teaser}B-E).
This section introduces the visualizations and interactions in the order our workflow progresses.
Experts can start using the system by specifying the dataset, the \cP{player} of interest, and the \cO{opponents} on the control bar.
The system spends seconds filtering the rallies played by the player against the opponents from the dataset and running the algorithm to obtain an initial tactic set.
Experts can explore these tactics in Projection View (Fig. \ref{fig:teaser}E), where we reveal the similarity among tactics (\ref{task:overview}, \ref{task:similarity}), and Tactic View (Fig. \ref{fig:teaser}C), where we list all tactics (\ref{task:overview}).
When they find tactics that should be adjusted, they can give suggestions in written language in the Suggestion Panel (Fig. \ref{fig:teaser}B) and preview the results (\ref{task:adjust}).
Projection View and Tactic View will then enable a preview mode to compare the new tactics with the original ones and present data-driven evaluations of the adjustment (\ref{task:evaluation}).
During the entire workflow, experts can view the rallies involving a tactic of interest in Rally View (Fig. \ref{fig:teaser}D) for detailed exploration (\ref{task:sequence}).
The visual designs in each view are as follows.

\subsection{Projection View}

Projection View (Fig. \ref{fig:teaser}E) projects the tactics onto a 2-D plane, helping experts overview tactics and reveal their similarity (\ref{task:overview}, \ref{task:similarity}).

\textbf{Projection.}
Experts can get an overview of all tactics from the overall layout. To enable this broad understanding and get across the effects of adjustments, the projection algorithm is essential.
After each adjustment, the system needs to re-run the projection algorithm to account for new tactics, where the position changes reveal the correlation between the new tactics and the original ones. Expert expect other tactics to keep stationary to avoid visual distraction. While t-SNE is a useful projection algorithm for visualizing data \cite{van2008visualizing,gove2022new}, it is not suitable for our system, because the results are not stable after each run.

\begin{figure}[tb]
    \centering 
    \includegraphics[width=\columnwidth]{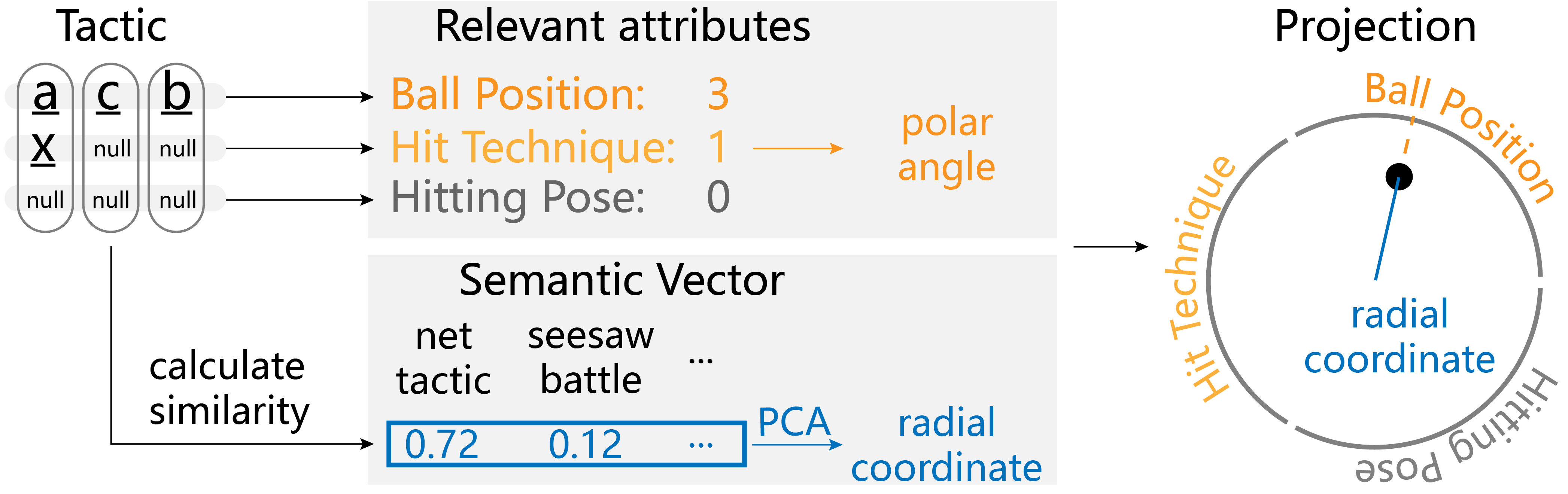}
    \vspace{-0.2in}
    \caption{An example that illustrates our proposed semantic-based projection, which mainly considers the two most relevant attributes and the basis tactics it resembles.}
    \label{fig:projection}
\end{figure}

Instead, we propose a semantic projection method based on a polar coordinate system, which can generate a fixed position for each tactic based on semantics (Fig. \ref{fig:projection}).
For each tactic, the polar angle encodes the top-two relevant attributes (i.e., the two attributes with the most non-null values) so that experts can analyze a specific tactic type by exploring the corresponding direction.
Meanwhile, we adopt PCA, which can generate fixed projection results, to project each tactic to a 1-D coordinate system as the \revise{SRC7,R4C3}{radius}{radial coordinate}.
However, tactics vary in length and contain categorical values that are difficult to quantize (e.g., the hitting technique), which cannot be used as inputs for PCA.
Thus, we characterize a tactic as a semantic vector based on several basis tactics, constructed in three steps:
1) We required experts to provide ten different typical tactics, such as seesaw battles and net tactics, to be used as basis tactics baked in our system.
2) We calculate the Levenshtein distance of the tactic at hand from each basis tactic to represent their similarity, following TacticFlow.
3) We construct the vector with the similarities in order and normalize it.

\textbf{Encodings of each point.}
\revise{R3C7}{}{
Experts prefer to quickly find the tactics worthy of analyzing (i.e. with high \textit{frequency} or high \textit{tactical importance}) and then evaluate the effectiveness of these tactics (i.e., the \textit{win rate}).
All three variables are important, but encoding them simultaneously can lead to severe visual clutter, especially when there exist many points.
Thus, we encode either the frequency or the importance with the size of each point, allowing users to manually toggle in a settings panel.
The color from green to red encodes the win rate from 100\% to 0\%.}
When experts find a tactic of interest, they can turn to Tactic View for further exploration in two ways.
1) We display the ranking numbers of the top-ten tactics in Tactic View on the corresponding points.
2) Hovering over a point can highlight the tactic in Tactic View.

\subsection{Tactic View}

\begin{figure}[tb]
    \centering 
    \includegraphics[width=\columnwidth]{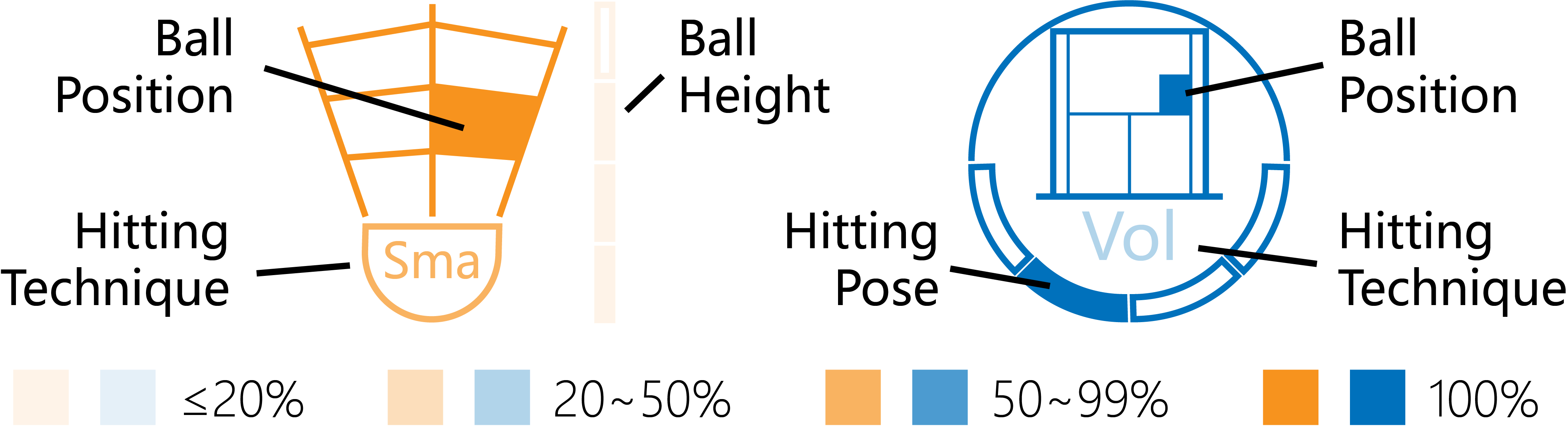}
    \vspace{-0.2in}
    \caption{The glyphs for badminton (left) and tennis (right), consisting of multiple components encoding detailed hit features. For each null value, we apply a multi-level opacity encoding for the frequency percentage of the value with highest frequency.}
    \vspace{-0.15in}
    \label{fig:glyph_design}
   \end{figure}

Tactic View (Fig. \ref{fig:teaser}C) lists all tactics (\ref{task:overview}). Each row represents a tactic and contains six columns as follows.

\textbf{Ranking Columns} (Freq., Win\%, Imp.).
Experts may start by ranking the list to find the most valuable tactics.
We use lineups \cite{gratzl2013lineup} to provide multiple rankings based on the frequency (Freq.), the win rate (Win\%), and the tactical importance (Imp.), visualized by the three bars.
Experts can select one of the three to satisfy different analysis needs.

\textbf{Tactic Column} (Tactic).
After selecting a suitable ranking, experts may scan the tactic list top-down and explore the tactics one by one.
\revise{R3C2.A}{}{Each hit in each tactic is encoded by a tailored glyph, which aggregates the corresponding hit in the raw rallies where this tactic was used, with the color encoding the player who hits the ball.}
The glyph is based on a metaphorical icon, such as the shuttlecock and the tennis ball (Fig. \ref{fig:glyph_design}), familiar and intuitive to experts \cite{wu2017ittvis,wu2020visual,wang2021tac,maguire2012taxonomy,ying2021glyphcreator,ying2022metaglyph}.
Each glyph encodes a multivariate hit with multiple non-overlapping components, each encoding a hit feature.
For example, in the tennis glyph in Fig. \ref{fig:glyph_design}, the blue block on the tennis court encodes the position where the ball bounces on the ground.
All detailed designs are in the appendix.

The main novelty of our glyphs lies in encoding null values with uncertainty, each of which indicates that there exists multiple possible values every time using the tactic.
For each null value, we first count the frequency of each possible value.
Then, we display the possible value with the most frequency in our glyph, where the opacity of the corresponding component varies on multiple levels based on the frequency percentage.
Such a design can help experts see how the tactic most often progresses, as well as how uncertain each value is.
Furthermore, experts can expand a tactic to find the two most possible values for each null, where the bars visualize the frequency percentage.

\textbf{Operation Columns} (No., Pref.).
When experts are satisfied with a tactic, they can click the favorite icon (Pref.) to fix it, avoiding future adjustments.
When experts find some tactics to adjust, they can select the tactics (No.) and give suggestions in the Suggestion Panel.

\subsection{Suggestion Panel}

\revise{SRC2,R1C2,R3C5,R4C1}{}{
Experts can give new suggestions in the top input box (\ref{task:adjust}) and view past suggestions in the list below (Fig. \ref{fig:teaser}B), where we allow experts to express their suggestions in natural language.
Natural language is one of the best methods to enable experts to interact with complex algorithms and models\cite{yu2019flowsense,shen2021towards,luo2021natural,gao2015datatone}.
Considering that experts are generally not familiar with math-based constraints, there is a steep learning curve for them to learn the constraint space, let alone to express their suggestions through constraints.
Thus, we implement a template-based NLP algorithm by NLTK in Python.
The templates are built based on the suggestions collected in the pre-study described in Sect. \ref{sec:prestudy}, such as ``$\langle$hit features$\rangle$ $\langle$is/are$\rangle$ important for $\langle$tactics$\rangle$'' (c5 in Fig. \ref{fig:constraints}).
After experts input a sentence in the top input box, the algorithm will try to match the sentence with each template, thereby finding the constraint type and extracting the parameters.}
To ensure the correct mapping, we display the mapped constraint and the parsed parameters and allow experts to modify them directly.
In addition, experts can undo the most recent adjustment and click on a suggestion in the history list below to view the corresponding historical data.

\subsection{Preview Mode}

Experts can preview the results of an adjustment after giving a suggestion and then determine whether to apply the adjustment (\ref{task:evaluation}), through the preview mode in Projection View (Fig. \ref{fig:teaser}E1) and Tactic View (Fig. \ref{fig:teaser}C1).
C1 and E1 demonstrate one example\revise{R1C3}{,}{} of splitting one tactic (with solid border) into two (with dashed borders).
The two new tactics have different colors, indicating different win rates.
Experts may regard this as a meaningful adjustment because the original combination of these two tactics had hidden their differences.
They may turn to Tactic View for detailed exploration.
Tactic View moves the three tactics to the top for comparison, with icons \cL{-} and \cW{+} representing the original tactic and the new tactics, respectively.
To reveal their differences in detail, experts can compare the glyphs and even click on each tactic to observe the related rallies in Rally View.

\subsection{Rally View}

Rally View (Fig. \ref{fig:teaser}D) provides the most detailed type of information --- the rallies related to a tactic (\ref{task:sequence}).
This view consists of three sub-views that may be useful for tactical analysis, explained below from top to bottom:
1) The stacked bar chart visualizes \textbf{the hit indexes of the tactic}, helping experts know where the tactic is and should be used.
The x-axis and y-axis represent the hit index and the frequency, respectively.
The green and red bars represent the number of wins and losses, respectively, when the tactic starts with the corresponding hit.
2) The two rally lists display \textbf{the detailed hit features of each rally}. One list shows winning rallies and the other shows losing ones.
Each row shows one rally, and each circle represents a hit, where solid circles encode the hits involving the tactic being explored.
Experts can further explore the detailed hit features of each hit by clicking the rally.
3) Experts prefer to watch \textbf{the video fragment of each rally} to find details not recorded in the data, such as players' facial expressions.

\section{Evaluation}

\begin{figure*}[tb]
    \centering 
    \includegraphics[width=\linewidth]{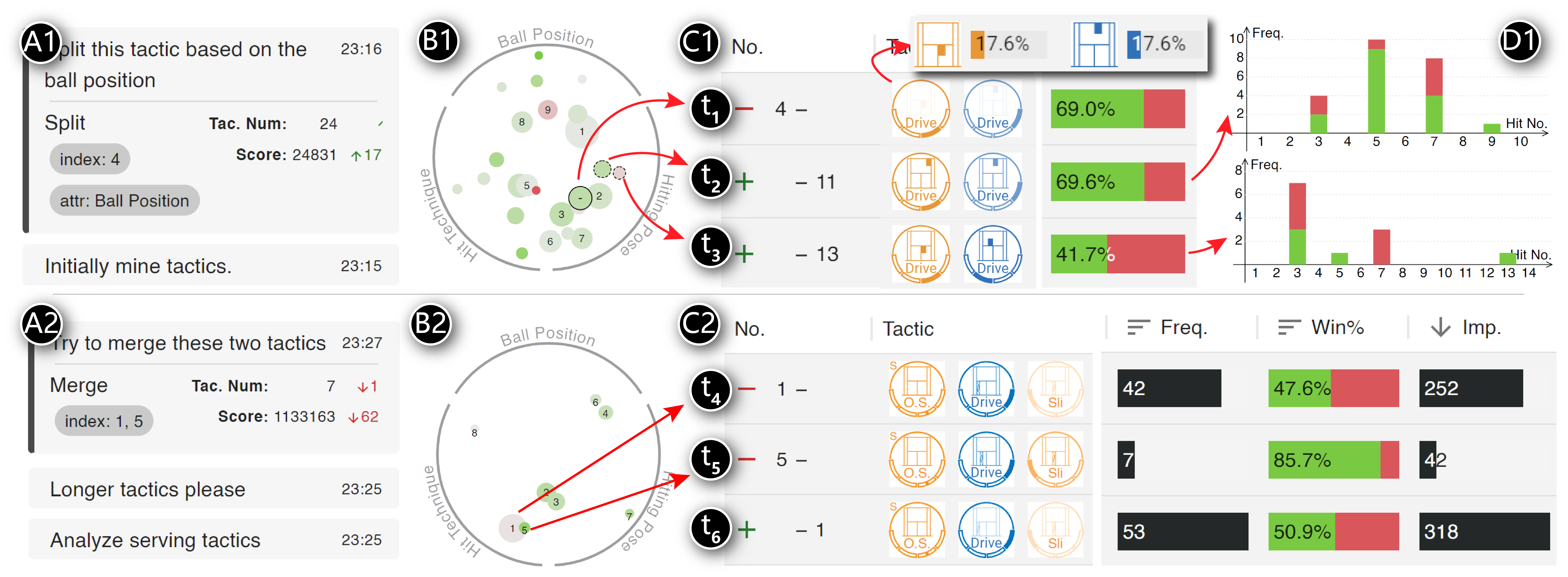}
    \vspace{-0.2in}
    \caption{Screenshots of the tennis case, where experts split a seesaw-battle tactic (A1-D1) and merged two serving tactics (A2-C2). We present the screenshots of Suggestion Panel (A1, A2), Projection View (B1, B2), Tactic View (C1, C2), and the stacked bar chart in Rally View (D1).}
    \vspace{-0.15in}
    \label{fig:case1}
\end{figure*}

\revise{SRC4,R2C4}{}{We evaluate our algorithm's performance and our system's usefulness through two quantitative experiments and two case studies, respectively.}

\subsection{Algorithm Evaluation}

\subsubsection{Tactic Quality}

\label{sec:evaluation}

\revise{SRC4,R2C4}{}{We first evaluate the quality of the adjusted tactics. We invited four experts (E1-E4) to conduct this experiment, testing whether our system could discover the expected tactics pre-set by experts.}

\revise{SRC4,R2C4}{}{\textbf{Setup}.
We first required the experts to provide 100 tactics as benchmarks.
Each expert was required to select five matches they were familiar with and manually list five tactics for each match based on their knowledge about the specific players.
After experts repeatedly watched the game video and ensured the correctness of the listed tactics, we regarded these tactics as the benchmark tactics.
}

\revise{SRC4,R2C4}{}{\textbf{Process}.
We asked the experts to use our system to mine tactics, from the matches selected by other experts studying the same sport (e.g., both E1 and E2 studies in tennis, therefore E1 needed to mine tactics from the matches selected by E2), ensuring that the expert did not know the benchmark tactics in advance.
After experts finished mining, we recorded the number of benchmark tactics discovered by our system and the number of constraints needed to adjust certain initially mined tactics to each benchmark tactic (\textit{note: if there existed a mined tactic that had at most one value different from a benchmark tactic, we regarded this benchmark tactic to be discovered by our system}).
}

\revise{SRC4,R2C4}{}{\textbf{Results}. The results were analyzed from three angles as follows.
\begin{itemize}[nosep,leftmargin=*]
    \item \textit{Effectiveness}: Following experts' suggestions, our system discovered 94 tactics from a total of 100 benchmark tactics, which proved the effectiveness of our algorithm. The other 6 tactics were all used in the tiebreaker (i.e., the last game to decide who was the winner of the match), crucial (hence found by experts as the benchmark tactics) but not frequent (hence not captured by our system).
    \item \textit{Efficiency}: For each captured benchmark tactic, experts applied 0.38 global constraints and 3.11 local ones on average, which proved that our system can quickly meet experts' expectations.
    \item \textit{Necessity}: Only 3 out of benchmark tactics were directly mined by the initial generation process without any constraints, which proved the necessity of interactive pattern mining.
\end{itemize}
}

\subsubsection{Runtime}

\begin{table}[tb]
    \caption{The results of the quantitative experiments.
    We generated 6 synthetic datasets ($D_1$ to $D_6$), varying in the number of sequences ($S$), the length of each sequence ($|s_i|$), the number of features in one hit ($k$), the number of tactics ($|T|$), and the number of options for each hit feature ($|V|$).
    For each dataset, we present the average runtime of the initial mining ($t_i$) and applying global ($avg.$ $t_g$) and local constraints ($avg.$ $t_l$).}
    \label{tab:experiments}
    \scriptsize%
      \centering%
    \begin{tabu}{*{9}{r}}
    \toprule

    \multicolumn{6}{l}{Datasets}&\multicolumn{3}{l}{Results} \\

    \cmidrule(lr){1-6}\cmidrule(lr){7-9}

    $id$ & $|S|$ & $|s_i|$ & $k$ & $|T|$ & $|V|$ & $t_i$(s) & $avg.$ $t_g$(s) & $avg.$ $t_l$(s) \\

    \midrule

    $D_1$ & 500 & 10 & 3 & 25 & 10 & 51.4 & 50.6 & 0.02\\

    $D_2$ & \textbf{700} & 10 & 3 & 25 & 10 & 97.3 & 97.1 & 0.03\\

    $D_3$ & 500 & \textbf{20} & 3 & 25 & 10 & 73.0 & 73.4 & 0.05\\

    $D_4$ & 500 & 10 & \textbf{5} & 25 & 10 & 80.2 & 80.1 & 0.03\\

    $D_5$ & 500 & 10 & 3 & \textbf{50} & 10 & 74.5 & 73.3 & 0.01\\

    $D_6$ & 500 & 10 & 3 & 25 & \textbf{20} & 28.7 & 28.9 & 0.02\\

    \bottomrule
    \end{tabu}%
    \vspace{-0.2in}
  \end{table}

\revise{SRC4,R2C4}{}{Given that experts need to continually adjust tactics, we further evaluated our algorithm's runtime, to ensure that experts can interact with the system smoothly.}
We ran our algorithm on multi-scaled synthetic datasets and recorded the runtimes required to mine the initial tactic set, process global constraints, and process local ones (Table \ref{tab:experiments}).

\textbf{Setup}.
We generated six synthetic datasets varying over five parameters, and generated random sequences and random tactics (each with three hits and seven non-null values) for each dataset.
Each tactic was embeded into 10\% of the sequences, therefore we can regard the generated tactic set as an appropriate best model for the generated dataset.
For each tactic set, we generated 4 global constraints (one for each type, preventing conflicts like expecting long tactics and expecting short ones) and 30 local ones (five for each type).

\textbf{Process}.
We applied the constraints for each dataset in random order, as experts would \revise{R1C3}{}{do} in an actual analysis scenario.
The time necessary to process each constraint is recorded for analysis.

\textbf{Results}.
The runtime shows that our algorithm can support smooth interactions.
For each dataset, the runtime required to process global constraints was similar to the time needed for initial mining (about one minute), which is acceptable for experts.
The runtime required to process local constraints was almost negligible because our generator can accurately generate only a few candidates expected by experts.

\subsection{Case Studies}

We invited four out of the five experts to conduct two case studies each in tennis (E1, E2) and badminton (E3, E4).

\subsubsection{Tennis}

The initial tactic set contained 23 tactics from 604 rallies played by Djokovic against Thiem (334), Nadal (143), and Pouille (127) since 2019.
\revise{SRC7,R4C3}{}{The tactics involved three hit features (encoded by the glyph on the right of Fig. \ref{fig:glyph_design}) — the ball position (i.e., the position where the ball bounced on the court, encoded by the court at center), the hitting technique (which is displayed by the abbreviation text), and the hitting pose (i.e., four types of hitting pose encoded by the donut below).}
The analysis process occurred as follows.

\textbf{Splitting a seesaw-battle tactic based on the ball position.}
Experts were first attracted by the text ``\revise{R4C5.A}{Dri.}{Drive}'' in the top tactic in Tactic View ($t_1$ in Fig. \ref{fig:case1}C1), which represents the hit technique \textit{drive} --- \ec{Texts are easier to understand than visualizations.}
Tactic $t_1$ involved both players consecutively driving the ball, indicating a seesaw-battle tactic.
Based on their knowledge, experts thought that the ball position decided the outcome of a seesaw battle.
However, they found a low opacity of the courts in the glyphs, indicating high uncertainty around ball positions.
Experts further expanded the tactic and found that the ball position with the highest frequency was used in only 17.6\% of rallies.
E1 said that, \ec{the tactic is so abstract that I could not obtain insights from it.}
Thus, they selected the tactic and suggested \ec{splitting the tactic based on the ball positions} in Suggestion Panel (Fig. \ref{fig:case1}A1).

\textbf{Analyzing the split tactics.}
Then, experts previewed the results and compared the two new tactics  ($t_2$ and $t_3$) with the original ones.
They noticed that, in Projection View (Fig. \ref{fig:case1}B1), the new points were closer to the sector of ball position, indicating that new tactics contained more non-null values of ball position.
E2 praised the design, saying \ec{the projection helps me track the changes.}
Experts found that, although $t_2$ and $t_3$ were similar, they had different win rates --- hitting the ball to the baseline ($t_2$) tended to bring more wins for Djokovic than hitting it to the service line ($t_3$).
To uncover the reason, experts further explored $t_2$ and $t_3$ in Rally View (Fig. \ref{fig:case1}D1).
The bar chart showed that Djokovic preferred to use $t_2$ after the fifth hit, but to use $t_3$ at the third hit, which is the first attack after service.
Experts thought that the opponents might return the ball weakly at the second hit, making it difficult for Djokovic to return powerfully at the third hit, \ec{like a ball falling on the ground, the faster it hits the ground, the faster it bounces.}
However, his opponents can return powerfully at the fourth hit in an attempt to gain the advantage.
By viewing the rallies and watching the videos, experts confirmed their thoughts and then applied the adjustment.

\textbf{Analyzing serving tactics.}
After splitting the seesaw-battle tactic, a serving tactic became the top tactic in Tactic View.
\revise{R4C2}{Due to the significant impact of serving tactics on the outcomes}{Because Djokovic excels at serving tactics}, experts suggested \ec{analyzing serving tactics} in Suggestion Panel (Fig. \ref{fig:case1}A2).
However, experts were not satisfied with the new tactic set because almost all serving tactics contained only two hits.
Experts further suggested \ec{analyzing longer tactics} and obtained some three-hit serving tactics.
E2 re-ranked the tactics in Tactic View (Fig. \ref{fig:case1}C2) based on the tactical importance and commented that, \ec{the tactical importance is more valuable than the frequency as frequent tactics are usually short and lack analytical value.}

\textbf{Merging two similar serving tactics.}
When analyzing the most important tactics recommended by our system ($t_4$), experts found that the point representing this tactic in the Projection View was overlapped with a point representing another tactic ($t_5$), indicating the two tactics were similar (Fig. \ref{fig:case1}B2).
Experts expected to merge them to form one tactic.
However, the point representing $t_4$ was nearly gray, while the point for $t_5$ was green, indicating different outcomes and leaving experts hesitant.
Given that they would be able to undo the adjustment, experts decided to try.
E1 commented that, \ec{trial-and-error tactical analysis is essential.}
Experts found that our system outputted a new tactic set with a score similar to the original one, which supported the adjustment and made experts confident (Fig. \ref{fig:case1}A2).
In Tactic View, experts noticed that the new tactic ($t_6$) had a similar win rate to $t_4$, with little impact from $t_5$.
Experts guessed that $t_5$ had a low frequency, which may lead to a misleading win rate --- \ec{Djokovic might win these rallies by chance}.
Experts applied \revise{R4C5.D}{the}{this merging} adjustment and obtained a more accurate estimate of the win rate for this serving tactic.

The two experts mainly gave positive comments on the interactive analysis workflow in the following interview.
\revise{R4C5}{}{But E2 mentioned a limitation in fine-tuning tactics -- \ec{The system accurately mined the tactics I wanted, but a significant adjustment might require numerous commands. It can be helpful to control the extent of an adjustment.}}

\subsubsection{Badminton}

\begin{figure}[tb]
    \centering 
    \includegraphics[width=\columnwidth]{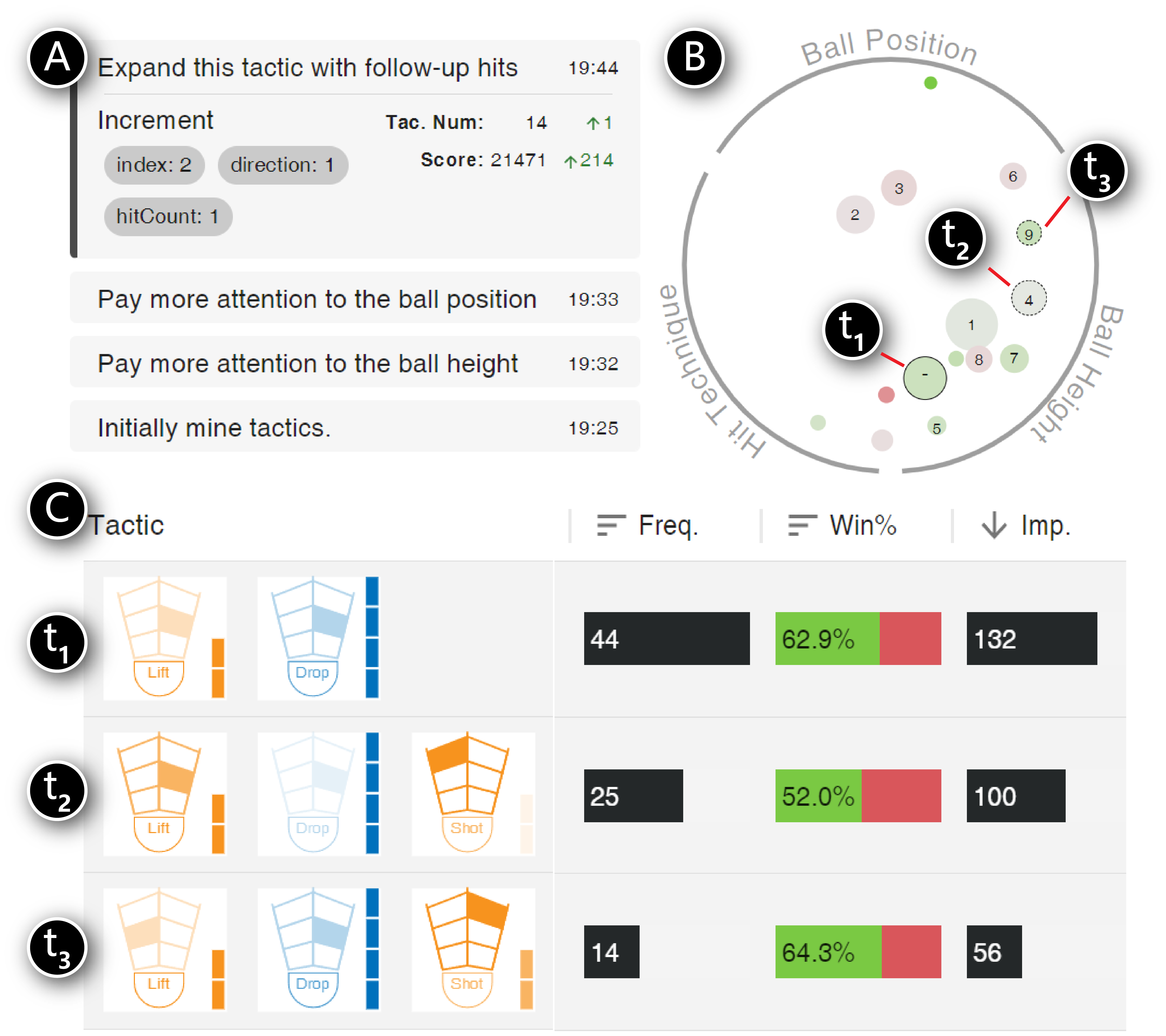}
    \caption{Case in badminton. Experts gave suggestions for adjusting tactics (A). When experts suggested expanding the tactic, the system enabled the preview mode of Projection View (B) and Tactic View (C).}
    \vspace{-0.15in}
    \label{fig:case2}
\end{figure}

The dataset contained 377 rallies played by Momota Kento against Viktor Axelsen (147), Anders Antonson (104), Srikanth Kidambi (63), and Yuqi Shi (63) since 2019.
\revise{SRC7,R4C3}{}{Experts mainly considered three hit features (encoded by the glyph on the left of Fig. \ref{fig:glyph_design}) — the 2-D ball position (i.e., the position where the player hit the shuttle, encoded by the top-left court), the ball height (i.e., a four-level quantization of the maximum ball height encoded by the right bars), and the hitting technique (which is displayed by the bottom-left abbreviation text).}
The algorithm mined 13 initial tactics from these rallies for experts to analyze.
The exploration process was as follows.

\textbf{Finding a tactic inconsistent with experts' knowledge.}
Tactical analysis in badminton usually focuses on both the 2-D ball position and the ball height.
Hitting techniques are also named based on the ball position (e.g., \textit{lift} is a technique hitting the ball from a low position at the player's backcourt to a high position at the opponents' backcourt).
Thus, experts first suggested paying more attention to the ball position and the ball height, leading to two constraints on the importance of these two attributes (Fig. \ref{fig:case2}A).
Then, experts started from Projection View to explore the tactics in the sector of the ball height, i.e., the tactics most relevant to the ball height (Fig. \ref{fig:case2}B).
E4 praised the semantic-based projection, \ec{different directions corresponding to different tactics is intuitive to me, helping me start analyzing.}
Experts were quickly attracted by a point (with a black border in Fig. \ref{fig:case2}), which was large (high importance) and green (high win rate).
Turning to Tactic View, experts further explored the tactic with two hits (t1 in Fig. \ref{fig:case2}C) --- Kento first lifted the shuttlecock, and his opponents then dropped the shuttlecock (i.e., an offensive technique that involves hitting the ball from a high position to a low position in the opponents' court).
Experts were confused by the tactic because \ec{Kento gave his opponents a chance to attack but finally won many points.}

\textbf{Exploring the tactic to find the reason.}
\revise{R4C2}{}{Because Kento did not mainly rely on lift technique to score points}, experts thought that the tactic must have further follow-up development and suggested \ec{expanding the tactic with follow-up hits} (Fig. \ref{fig:case2}A).
Our system presented two new tactics with three hits ($t_2$ and $t_3$ in Fig. \ref{fig:case2}C).
In the third hit of two of the tactics, Kento hit the shuttlecock at a low height and to the opponents' backcourt (to the left in $t_2$ and the right in $t_3$).
However, the text ``Shot'' in the two tactics (i.e., a technique that involves hitting the ball a little higher than the net) attracted experts, as it indicated Kento's counterattack.
Experts continued to watch the video and found that Kento was familiar with the tactic, indicating that he might train for it. 
He used a lift for the first hit to lure his opponents to drop the shuttlecock.
Then, relying on the quick reaction developed through training, he can counterattack with the shot technique, often throwing off his opponents and earning him points. 
Experts said that \ec{the system helped us find Kento's secrets behind the tactic.}

The two experts gave positive comments on our system in the following interview.
E3 said that, \ec{it is like I am communicating with the computer and teaching it to find tactics.}
\revise{R4C5}{}{E4 suggested supporting tactic searching -- \ec{We need to search for some crucial tactics that are not mined by the system due to the low frequency.}}

\section{Discussion}

\textbf{Scalability.}
We discuss the scalability of the system's two main views --- Projection View and Tactic View.
First, \revise{R3C2.B}{}{
the projection view cannot precisely encode multiple related hit features of a hit.
Although we can scale the projection view to more hit features by simply splitting the polar coordinates into more sections, users can hardly judge which hit feature is the second most important. 
But the projection view can accurately encode the most important parameters, which can help experts explore a certain type of tactics, crucial and sufficient for experts.}

Second, when many hit features exist in one hit, we cannot always draw all of the components in a single glyph in the Tactic View.
In the future, experts may expect to analyze other hit features that are not presented in the current dataset, such as the speed of the ball.
Drawing many hit features in one glyph can lead to high visual clutter.
A possible solution is to encode the hit features that experts most care about in the glyph and display others in tooltips.

\textbf{Generalizability.}
Recently, we have witnessed that many domains like sports and EHR recorded and analyzed multivariate event sequences rather than only analyzing the event type and the timestamps.
\revise{R4C8.F}{}{Our work can inspire researchers in other domains to study interactive pattern mining and obtain better mining results. Meanwhile, our workflow with close collaboration with domain experts can be instructive.}

\textbf{Design Implications.}
We summarize several findings made during our study, which may be relevant to interactive visual analytics in other domains. 
1) Comparative analysis is crucial for interactive tactical analysis, during the process of which data-driven metrics play an essential role to provide hard evidence about new results, helping users to better trust the system. 
2) Semantics could help domain experts with little visualization experience explore a projection view. If the position of the points presents semantic information, this could help experts recognize and identify each point from their domain perspective. 
3) The uncertainty arising from the interactive analysis process is an issue, especially when users interact with the system through natural language. When users do not understand how the system works, they may make intractable suggestions, preventing the system from producing results that meet users' expectations. We tried to address this issue by exposing how user suggestions are translated to constraints step by step. Further research is required to better address the problem. 

\textbf{Limitations and future work.}
\revise{SRC2,R2C2,R4C4}{}{Although we summarized nine constraints and hundreds of natural language templates through the pre-study to support interactive tactic mining, we still found them limited the diversity of tactics in practice. For example, players prefer to hide some tactics as weapons used in tiebreakers, which are crucial (hence requiring analysis) but infrequent (hence tricky for algorithms to discover). Without similar tactics to modify, experts can hardly obtain these tactics in the tactic set for further analysis. Natural language can be hardly used to precisely define a complex tactic with multiple detailed hit features due to the limited natural language templates and the ambiguity of natural language. Compared with NL-based interactions, Eventpad \cite{cappers2018eventpad} proposed efficient and intuitive interaction widgets to help users construct multivariate patterns for further queries based on regular expressions. Our future work includes mining these crucial but infrequent tactics and integrating a pattern searching function.}

\section{Conclusion}

This work introduces a visual analytics system for interactive tactical analysis in racket sports, which allows experts to incorporate their knowledge into data mining algorithms to discover meaningful tactics.
We propose a constraint-based pattern mining algorithm that discovers an initial set of tactics and then supports adjustments to them by translating experts' written suggestions into further constraints.
We also propose a user interface through which experts can interact with the algorithm --- exploring tactics through metaphoric glyphs, giving suggestions, and verifying results through comparative visualizations.
A quantitative experiment on synthetic datasets shows that our system supports smooth interaction between experts and the algorithm.
Two real-world case studies demonstrate that our system can help experts use their knowledge to find meaningful tactics.
\end{spacing}

\vspace{0px}
\acknowledgments{
    The work was supported by NSFC (62072400) and the Collaborative Innovation Center of Artificial Intelligence by MOE and Zhejiang Provincial Government (ZJU).}

\bibliographystyle{abbrv-doi-hyperref-narrow}

\bibliography{template}
\end{document}